\documentclass[letterpaper]{article}

\usepackage[T1]{fontenc}

\usepackage{geometry}
\usepackage{setspace}

\usepackage{achemso} 
\setkeys{acs}{articletitle = true}
\usepackage{graphicx}
\usepackage{float}
\newfloat{scheme}{htbp}{los}
\floatname{scheme}{Scheme}
\floatname{chart}{Chart}
\newfloat{graph}{htbp}{loh}
\usepackage{caption}

\usepackage{amsmath,amssymb} 
\newcommand{\angstrom}{\textup{\AA}}

\usepackage{authblk}
\author[]{Tom Rosenstein}
\author[]{Philipp Zolthoff}
\author[]{Jan Kierfeld*}
\author[]{Matthias F. Schneider*}
\affil[]{Department of Physics, TU Dortmund University, 44227 Dortmund, NRW, Germany}

\title{Generation of two-dimensional pulses in lipid monolayers by rapid photoswitching}
\date{*Email: jan.kierfeld@tu-dortmund.de, matthias-f.schneider@tu-dortmund.de}

\begin{document}

\maketitle

\begin{abstract}
    We study pressure pulse generation and propagation in lipid monolayers by an experimental approach employing rapid photoisomerization of photoswitchable lipids (azoPC). This allows us to generate longitudinal surface pressure pulses 
    by optical flash excitation in both free and constrained layer geometries.
    We compare the observed pulse shapes with a theoretical approach based on a nonlinear fractional wave equation for a surface displacement field, where a fractional time derivative term captures
  the hydrodynamics of the monolayer subphase. We explore channel geometries of different lengths and widths and find quantitative agreement between theory and experiment regarding pulse speed and pulse shapes. For narrow channels, we employ a one-dimensional version of the fractional wave equation 
  to study pulse propagation without any fit parameters by using the pressure signal at a 
  close pressure sensor as boundary condition to predict the pressure signal at a second far sensor.  
    A full two-dimensional description can capture all effects arising from the channel geometry for wider channels using one common set of fit parameters for the pulse excitation that can be applied to all geometries. 
    The nonlinearity in the fractional wave equation plays no role in  explaining the observed pulse shapes because pulse amplitudes generated by azoPC photoswitching remain very small.
\end{abstract}

\section{Introduction}

Hydrated lipid membrane interfaces are ubiquitous in cells; plasma membranes form the outer boundary of cells and internal membranes compartmentalize cells.
Moreover membrane-bound vesicles are essential for intra- and intercellular transport.
Membranes are  selectively permeable and can embed a variety of proteins and protein complexes at high concentrations. 
As two-dimensional fluids with compression and bending moduli they can potentially support mechanical waves and pulses, which are a consequence of momentum conservation and the existence of internal restoring forces giving rise to compression and bending resistance.
Local perturbation can trigger chemical or electrical waves and pulses often supported or mediated through proteins.
One prominent example of an electric wave is the action potential pulse in nerve cells.
Various studies have shown swelling as well as the release and absorption of heat during action potentials, leading to the hypothesis that action potentials also involve an adiabatic mechanical solitary wave, i.e., an acoustic pulse \cite{Heimburg2005}.
Coupling to the surrounding water as well as membrane viscosity gives rise to dissipation and damping of pulses.

Monolayers of lipids at the air-water interface are a simpler model system to study important aspects of membranes. 
In particular, they allow for controlled mechanical experiments, because their mechanical properties are well characterized via pressure-area isotherms, methods for  mechanical excitation exist, and local pressure measurements by Wilhelmy plates are possible \cite{Lucassen1968a,Griesbauer2009,Griesbauer2012,Shrivastava2013,Shrivastava2014}.   
Various excitation methods have already been explored in the past, such as  local expansion or condensation by drops of solvent or acid \cite{Griesbauer2012,Shrivastava2013}, electro-mechanical excitation \cite{Griesbauer2009}, short bursts of gas \cite{Fichtl2016,Fichtl2016b},  or a metal cantilever in the monolayer \cite{Shrivastava2014}.
All of these methods pose two major problems: excitation is not confined to the plane and they are not easily applicable to living systems. 

\begin{figure}
  \centering
  \includegraphics[width = 0.6\textwidth]{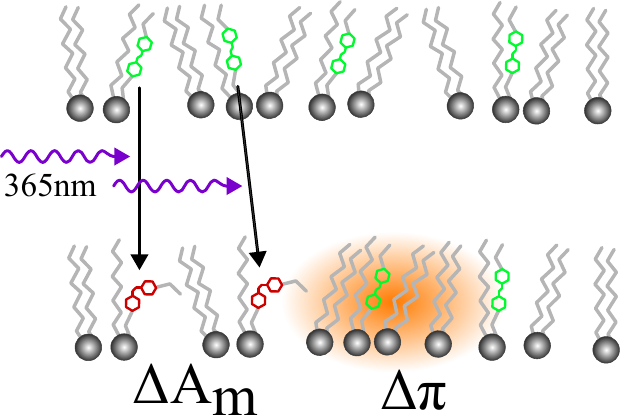}
  \caption{Schematic depicting how rapid photoisomerization of azoPC in monolayers from trans (green) to cis (red) via UV light generates a compression pulse ($\Delta\pi$) trough a local increase in molecular Area ($\Delta A_m$).}
  \label{fgr:intro}
\end{figure}

In this work we present an optical method of pulse generation, inspired by the work of Suzuki, Möbius and Ahuja \cite{Suzuki1986}, that solves these issues and present results on model membranes as a basis for further research.
This is achieved by incorporating azoPC, a phosphatidylcholine with an azobenzene moiety in one of its side chains into lipid monolayers, see Fig.\ \ref{fgr:intro}
\cite{Pritzl2025}.
Azobenzene exists in either the trans or cis isomer and can optically switch between both isomeric states.
The trans to cis transition is induced with UV light, while the cis to trans transition happens under irradiation with visible light or thermally \cite{Bandara2012}.
The latter, however, happens on the order of days at room temperature \cite{Liu1992}, thus yielding two very stable states to switch between.
Since the cis-isomer occupies a larger area per molecule than the trans-isomer, photoisomerization leads to a state change, which, in the isochoric case, manifests as a change in lateral pressure. \cite{Iwamoto1994, Warias2023}

Monolayers  that are mechanically excited by an in-plane force can support longitudinal mechanical waves because of their compression modulus.  
Marangoni forces from surface tension variations due to lipid density gradients give also rise to fluid flows in the aqueous subphase.
Both viscous dissipation in the subphase as well as the intrinsic two-dimensional monolayer viscosity will give rise to damping.  
It is known since the pioneering work of Lucassen  \cite{Lucassen1968a,Lucassen1968b}  that the monolayer can support a longitudinal "Lucassen wave" with a dispersion relation that is characteristic for the presence of these damping effects. 
Recently, important theoretical progress has been achieved by Netz and coworkers via a description of these surface waves by a fractional wave equation \cite{Kappler2015,Kappler2017, Zendehroud2022,Simon2019}, where dissipation in the subphase is captured by a fractional time derivative of the lipid displacement field. 
The propagation speed of pulses can be obtained from the dispersion relation and has been shown to agree well with experimental results \cite{Griesbauer2012,Kappler2017}. 
Many aspects of a detailed and quantitative comparison between experiment and the theoretical description via a fractional wave equation comparison are, however,  still missing and will be provided in the present work. 
In particular, we will present a detailed comparison of pressure pulse shapes, essentially without any fit parameters for monolayer or subphase properties. 
This comparison will also include the actual two-dimensional channel geometry of the experimental setup and explore different channel geometries.
We will also address the question whether nonlinearities in the fractional wave equation from a density-dependent compression modulus are relevant to describe the experiments.   

\section{Materials and Methods}

\subsection{Experimental setup}

\begin{figure}
  \centering
  \includegraphics[width = 0.8\textwidth]{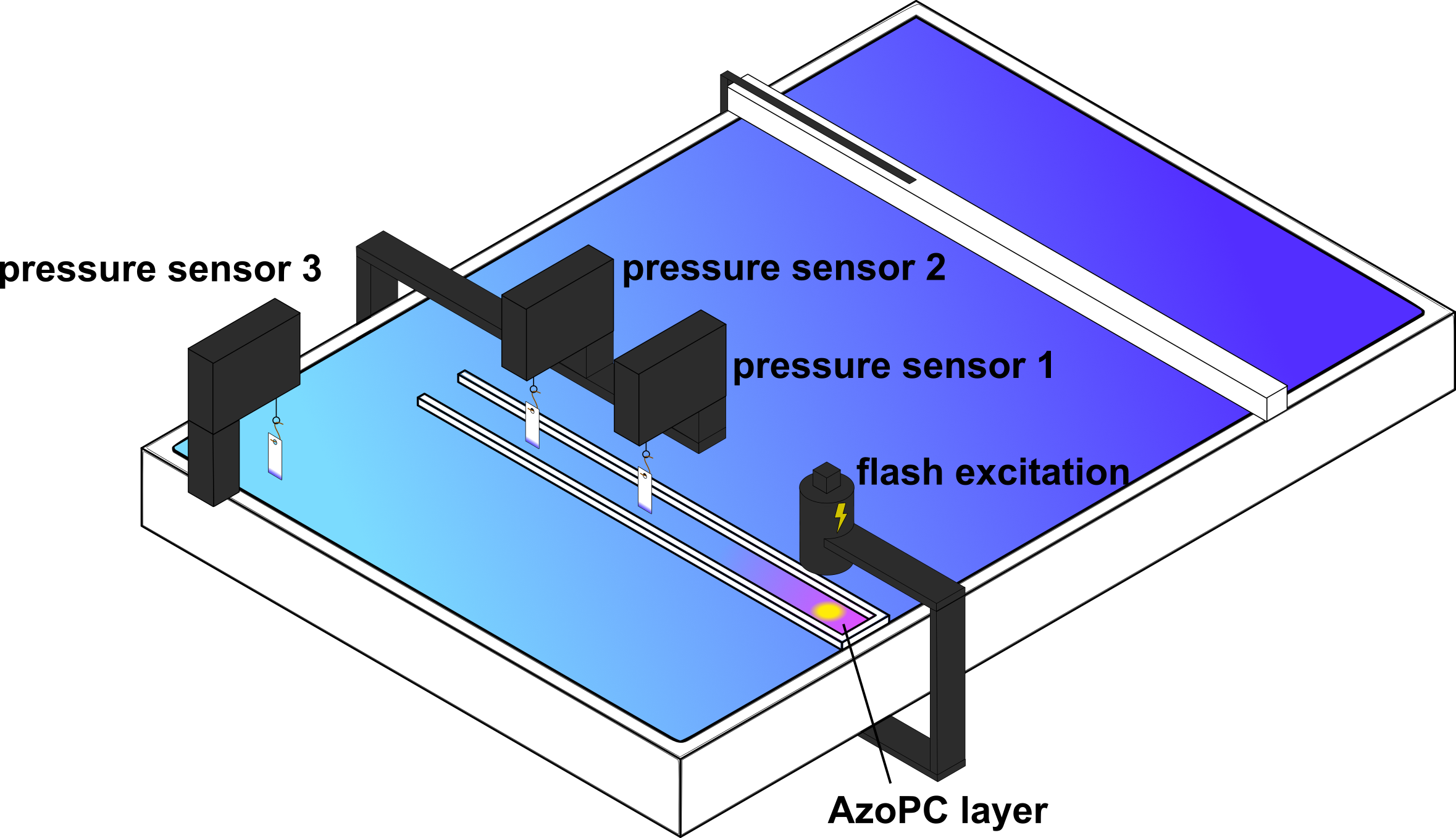}
  \caption{Schematic of the experimental setup.. Pressure sensors 1 and 2 are used to detect the pulses, while sensor 3 reports the overall state of the layer. In order to increase pulse amplitudes, teflon channels of different sizes can be placed in the trough as shown. Not shown in this illustration are 365\,nm and 455\,nm LEDs used to control the isomerization state of the monolayer as well as a heating bath connected to the trough for temperature control.}
  \label{fgr:exp}
\end{figure}

The general experimental setup is shown in Fig.\ \ref{fgr:exp}.
Phospholipid monolayers are prepared on a Langmuir-Blodgett trough by titrating a small amount of a given lipid solution onto the air-water interface.
Lateral pressure is measured at three different positions: Two pressure sensors (1 and 2) close to the area of excitation which are sampled at 1 kHz for a high temporal resolution. The third pressure sensor is further away and sampled at 10 Hz to monitor the state of the monolayer.
The trough is held at a constant temperature of $20\pm 0.5 {\rm ^\circ C}$ by a heating bath and the thermodynamic state of the interface is controlled by a movable polyoxymethylene (POM) barrier.
Optical Excitation is achieved by a customized commercial xenon flash (Metz Mecablitz 34, Flash time 4ms).
Diffusors and filters have been removed and replaced by an aspheric collimating lens and optical tubing which allows the use of different filters. A 245-400nm colored glass bandpass filter was used for the trans-to-cis excitation, while a 650nm shortpass (transmission region 400-640nm) filter was used for cis-to-trans excitation. Additionally 365nm and 455nm high power LEDs were employed to tune the trans/cis ratio of the monolayer.
In order to achieve higher pulse intensities different types of    
polytetrafluoroethylene (PTFE) channels can be placed inside the trough.
The chosen lipid solution is a 8:2 (molar) mixture of dipalmitoylphosphatidylcholine (DPPC) and azoPC solved in chloroform. Lipids were purchased from Avanti Polar Lipids (Alabaster, AL).

\subsection{Theory  and simulation}

Both the bulk subphase and the interfacial monolayer possess mechanical properties that must be incorporated into a unified theoretical framework\cite{Zendehroud2022}.
Fluctuation wave lengths are sufficiently small that we can linearize the Navier-Stokes equation 
for the velocity field $\mathbf{v}(\mathbf{r},t)$ of  the viscous subphase, 
\begin{equation}
    \rho \partial_t \mathbf{v} = \eta \partial_z^2 \mathbf{v}, 
    \label{eq:NS}
\end{equation}
where we also assumed that variations in the z-direction perpendicular to the interface in the xy-plane are dominant, $\mathbf{v}=\mathbf{v}(z,t)$; $\eta$ and $\rho$ are viscosity and density of the subphase.
The subphase can also be assumed to be incompressible, $\mathbf{\nabla}\cdot\mathbf{v}=0$.
The two-dimensional (2D) monolayer in the xy plane is modeled by a constitutive stress-strain relation 
that includes both elasticity and viscosity. 
The deformation state of an infinite  monolayer in the xy plane with a wave excitation propagating 
in x-direction 
is described  in terms of the tangential 
surface displacement field $\mathbf{u}(x,y,t)$ of the monolayer (with $\partial_t \mathbf{u}$ as surface velocity at the monolayer).
Viscous subphase and monolayer are coupled by continuity of tangential velocities at the interface (i.e., no-slip boundary conditions $\partial_t \mathbf{u}= \mathbf{v}(z=0)$) and interfacial stress equilibrium.
This leads to a subphase velocity field  that is largely tangential
apart from small normal components  due to incompressibility.

\subsubsection{One-dimensional (1D) model}

In the 1D model 
we assume that  the wave excitation is translationally invariant in y-direction and 
both the monolayer displacement field $\mathbf{u}=u(x,t)\mathbf{e}_x$ and the fluid 
velocity field $\mathbf{v}\approx v_x(x,z,t)\mathbf{e}_x$ are directed in pulse propagation direction. 
The interfacial stress equilibrium includes
the viscous shear stresses $-\eta \partial_z v_x(z=0)$ exerted by the subphase  and  the internal stresses 
of the monolayer, which are 
$k_{2d}(a)\partial_x^2u$ from spatial variations of the lipid area $a= \bar{a}(1+\partial_x u)$
(with respect to an average area $\bar{a}$) 
and, thus, the lipid density and from viscous stresses  $\eta_{2d}   \partial_t  \partial_x^2 {u}$ within the monolayer
with the elastic modulus $k_{2d}$ and the surface viscosity $\eta_{2d}$ of the monolayer.
Collecting all stresses that act on the surface, 
one can formulate the Marangoni stress condition $\nabla \sigma = - \nabla \pi$, where 
$\nabla \sigma = k_{2d}(a) \partial_x^2 u + \eta_{2d} \partial_t \partial_x^2 u$. In addition, a pulse activating 
Marangoni stress can be added in the form of pressure gradient $- \nabla \pi_{act}$ as an alternative to using the boundary conditions to cause perturbations.   
 Eliminating the subphase velocity field yields a retarded friction term \cite{Torvik1984,Kappler2017} for the monolayer that can be written as a Caputo-type fractional  time derivative \cite{Mainardi2010} of order $3/2$ 
 \footnote{The Caputo-type fractional  time derivative of order $3/2$  is defined by
 ${D}_t^{3/2} {u}(t) \equiv \frac{1}{\Gamma(1/2)} \int_0^t d\tau (t-\tau)^{-1/2} u''(\tau)$, where $\Gamma(x)$ is the Gamma function.
  }
 of the surface displacement field  in the interfacial stress equilibrium, 
 $-\eta \partial_z {v}_x(z=0) = - \sqrt{\eta \rho} \; {D}_t^{3/2} {u}$. 
For completeness, we also 
retain the inertial term associated with the monolayer's areal mass density $\rho_{2d}$, 
although it is typically negligible compared to the inertia of the underlying water subphase. 
The resulting equation of motion for an infinite  homogeneous monolayer in the xy plane with a wave excitation in x-direction as described by the surface displacement field $u(x,t)$ is given
by the interfacial tangential stress equilibrium, \cite{Kappler2017}
\begin{equation}
\rho_{2d} \partial_t^2 {u}  + \sqrt{\eta \rho} \; {D}_t^{3/2} {u}  
= k_{2d}(a) \; \partial_x^2 {u} + \eta_{2d}   \partial_t  \partial_x^2 {u}  .
\label{eq:DGL}
\end{equation}
Quantities labeled by the subscript $2d$ refer to in-plane material parameters of the monolayer.
Although equation (\ref{eq:DGL}) is formally linear in $u(x,t)$, the material parameters of the monolayer, the elastic modulus $k_{2d}$ and the surface viscosity $\eta_{2d}$ may vary with the thermodynamic state of the layer \cite{Heimburg2018}.
We retain linear kinematics, i.e.\ strains  $\epsilon \sim \partial_x u$, 
but allow for state-dependent moduli $k_{2d}(a)$ through the measured isotherm, 
where $a=a_0(1+\partial_x u)$ is the local area per lipid (with an average area $a_0$). 
An area-dependent 
modulus  $k_{2d}(a)$ thus constitutes an additional dependence on strain $\partial_x u$ and 
gives rise to an effective nonlinearity in eq.\ \eqref{eq:DGL}
when the area per lipid $a$ changes due to compression or expansion during wave propagation.
To be specific for the experimental situation,  we extract the  compressional modulus 
$k_{2d}=k_{2d}(a)= - a d\pi/da$ from the \emph{measured} 
surface-pressure isotherm $\pi = \pi(a)$ from Fig.\ \ref{fgr:isotherm}A; the results are shown in Fig.\ \ref{fgr:isotherm}B. 
We note that an analogous state dependence could be introduced for the viscosity, in principle.
We also note that we use the isothermal compressional modulus, 
which is related via $k_{2d} = 1/\kappa_T$ to the isothermal compressibility,
when  we calculate $k_{2d}$ from the isotherms.
Then we assume that the monolayer can release  
heat to the environment during compression. 
For fast processes such as pulse propagation 
it might be more appropriate to use the somewhat smaller adiabatic compressibility $\kappa_S<\kappa_T$
 and a corresponding larger adiabatic compression modulus \cite{Griesbauer2009},

where the moduli are related by the thermodynamic relation $1/k_s = 1/k_T - {T\alpha^2}/{C_p}$ \cite{Landau1986}.
Here $k_s$ and $k_T$ denote the adiabatic and isothermal compressional moduli, respectively, $\alpha$ is the thermal expansion coefficient, and $C_p$ is the heat capacity at constant pressure.
Due to experimental constraints, the isothermal modulus is considerably more accessible, while remaining sufficiently close to its adiabatic counterpart for the present purpose. The deviation is usually small.
Absolute values (as opposed to excess values) for the heat capacity of lipid monolayers are scarce in the literature 
because they are difficult to determine in calorimetric measurements. Using  a value of $1600\,\mathrm{J/mol/K}$ for DPPC
\cite{Blume1983}  and a thermal expansion coefficient of 
$\alpha \sim 3 \times 10^{-3}\, \mathrm{1/K}$, we obtain $T\alpha^2/{C_p}  \sim 0.5\,\mathrm{m^2/J}$.
Compared to  typical values $k_T \sim 50\,\mathrm{mN/m}$ this corresponds to a $2-3\%$ 
difference between adiabatic and isothermal compressional moduli.

In the absence of nonlinearities, the equation of motion eq.\ \eqref{eq:DGL} leads to a
Lucassen-like dispersion relation 
\cite{Lucassen1968a,Kappler2015}
\begin{equation}
  k^2 = \frac{\rho_{2d}\omega^2 + \sqrt{i\eta\rho \omega^3}}{k_{2d} + i\eta_{2d}\omega}.
  \label{eq:disperion}
\end{equation}

On the linear level, the dispersion relation directly gives the scaling relation between 
 lateral deformation length scales $L \sim 1/k$ and the corresponding timescales 
$T \sim 1/\omega$ that balance linearly elastic forces $\sim k_{2d}{U}/{L^2}$,  
surface-viscous forces $\sim \eta_{2d} {U}/{T L^2}$, and subphase-viscous forces 
$\sim \sqrt{\eta\rho}{U}/{T^{3/2}}$ in the equation of motion eq.\ \eqref{eq:DGL} with 
a typical  amplitude $U$.
The amplitude $U$  will be determined by balancing with the  pulse activating 
Marangoni force $\sim {\Pi}/{L}$ with a typical pressure amplitude $\Pi$: 
$U/L \sim \Pi/k_{2d}$.
Nonlinearities become relevant when the nonlinear correction governs  the compression 
modulus $k_{2d}(a)\sim k_{2d}(a_0)(1+U/L)$, which happens for $U/L \sim 1$, i.e., 
if pulse displacement $U$ become comparable to pulse sizes $L$. 
Typically the inertial contribution associated with $\rho_{2d}$ is negligible and also the monolayer viscous forces
associated with $\eta_{2d}$
are small compared to hydrodynamic dissipation forces via the subphase.  
Then we obtain a  scaling relation  
\begin{equation}
T \sim \omega^{-1} \sim   \left(\frac{\sqrt{\eta\rho}L^2}{k_{2d}}\right)^{2/3}
\label{eq:scaling}
\end{equation}
for typical relaxational timescales after deformation over length scales $L$.

In a linear theory, pulses propagate with a phase velocity \cite{Griesbauer2012,Kappler2017}
\begin{equation}
    c =\frac{\omega}{{\rm Re}\,k} =  
    \frac{1}{\cos \left({\pi}/{8} \right)} \sqrt{k_{2d} \sqrt{\frac{\omega}{\eta  \rho}}}.
    \label{eq:phase}
\end{equation}
In particular we expect small phase velocities close to the LC/LE transition, where the monolayer  compressibility $\kappa = 1/k_{2d}$ is maximal and, thus, the compression modulus $k_{2d}$ minimal.

The excitation of the layer will 
be implemented either through prescribed boundary displacements or via an externally applied pressure.
In the former case, the displacement at the channel entrance can be taken directly from the data measured at the 
first sensor, enabling a data-driven propagation model from the first to the second sensor.

\subsubsection{Two-dimensional (2D) model}

In realistic channel geometries we do not have strict translational invariance in y-direction
perpendicular to the channel, and we need a full 2D model. 
In 2D, the monolayer is modeled as an isotropic viscoelastic continuum supporting both shear and dilatational (compressional) modes. The 1D evolution equation generalizes to a 2D equation
\begin{align}
    \rho_{2d}\ddot{\mathbf{u}} + \sqrt{\eta \rho}\; {D}_t^{3/2} \mathbf{u}
    =
    \mu_s \Delta \mathbf{u} + \left(\lambda_s + \mu_s \right)\nabla \left(\nabla \cdot \mathbf{u} \right)
    \nonumber\\
    +\;
    \eta_s \Delta \dot{\mathbf{u}} + \left(\eta_s + \zeta_s \right)\nabla \left(\nabla \cdot \dot{\mathbf{u}} \right)
    - \mathbf{\nabla} \pi_{act},
\label{eq:DGL2d}
\end{align}
where $\mathbf{u}=\mathbf{u}(x,y,t)$ is the 2D in-plane displacement field, $\lambda_s$ and $\mu_s$ are the 2D Lam\'e parameters (dilatational and shear elasticity), and $\zeta_s$ and $\eta_s$ denote the corresponding surface (bulk and shear) viscosities.

In general, shear components of lipid monolayers are often small compared with dilatational contributions over the relevant frequency range. \cite{Choi2011}
At small areas per lipid, corresponding to the LC regime, shear effects are more pronounced than in less ordered regimes, such as the LE phase. Shear effects may also become more relevant in more structured layers, for example when lipids are capable of interdigitating.

For the channel boundaries, we impose an impermeability condition by enforcing vanishing normal displacement, $\mathbf{u}\!\cdot\!\mathbf{n}=0$, while allowing free tangential slip (no constraint on $\mathbf{u}\!\cdot\!\mathbf{t}$).
They connect to the 1D framework eq.~\eqref{eq:DGL}, which is done in Fig.~\ref{fig:1d_results}B, by taking the purely longitudinal reduction of eq.\ \eqref{eq:DGL2d}, yielding the identifications
\begin{equation*}
k_{2d} = \lambda_s + 2\mu_s,
\qquad
\eta_{2d} = \zeta_s + 2\eta_s.
\end{equation*}
In order to model the experimental pulse excitation via a light cone, we add a localized, time-dependent forcing $\mathbf{f}(x,y,t)$. In practice, we take $\mathbf{f} = - \nabla \pi_{act}$, where $ \pi_{act} = g(x,y)f(t)$ to be the in-plane pressure generated by an external stimulus, modeled as a circular symmetric Gaussian spatial profile $g(x,y) =  \exp(-(x^2+y^2)/2\xi^2)$  of width $\xi$ and a sigmoidal temporal profile, with a time constant $\tau$ characterizing the  surface pressure change after isomerization and an amplitude $f_0$,
\begin{equation}
    f(t) =   
    \frac{f_0}{2} \left( 1 + \tanh \left( {\frac{t-t_1}{\tau}} \right) \right).
\label{eq:f_t}
\end{equation}

\subsubsection{Simulation}

For the numerical solution, all spatial and temporal derivatives in eq.\ \eqref{eq:DGL} in 1D or 
eq.\ \eqref{eq:DGL2d} in 2D are discretized using standard finite-difference schemes, while the fractional Caputo term is approximated using an L2 discretization. \cite{Li2011}

To model wave propagation in 1D independent of the details of the excitation process, we can employ  the experimentally measured signal $\pi_1(t) = \pi_0 + \Delta \pi_1(t)$ of the first sensor as a Neumann boundary condition for the 1D model \eqref{eq:DGL}. This boundary condition directly encodes the measured lateral pressure $\pi(0,t)$ at the channel entrance $x=0$. 
Inverting the signal $\pi(0,t)$ by using  the measured (trans) isotherm $\pi = \pi(a)$ from Fig.\ \ref{fgr:isotherm}A,
we obtain a signal $a(t)$ at the channel entrance $x=0$. Using the 
relation $a = a_0\!\left(1+\partial_x u\right)$, this can be converted into a time-dependent 
Neumann boundary condition $\partial_x u(0,t)$ for $u$ at the channel entrance $x=0$. 
Alternatively, we can include the excitation process explicitly into the 1D model \eqref{eq:DGL}  or the 2D model 
\eqref{eq:DGL2d} as additional forcing term $-\partial_x \pi_{\mathrm{act}}$ from 
an externally imposed pressure $\pi_{\mathrm{act}}$ 
that directly mimics the photosensitive response of the monolayer upon isomerization of the azobenzene units.
After numerical solution in 1D or 2D the resulting time-dependent 
displacement field is converted back into a lateral-pressure signal 
by using  first
$a = a_0\!\left(1+\partial_x u\right)$ and then the measured isotherm $\pi = \pi(a)$ 
from Fig.\ \ref{fgr:isotherm}. This enables a direct comparison to the experimental pressure signals.
Alternatively, we could use the  
monolayer stress-strain relation 
$-\pi = \sigma(\epsilon) = k_{2d}(a)\partial_x u + \eta_{2d}   \partial_t \partial_xu $ in order to include also (small) dynamic effects from the viscous stresses.
To account for the finite size of the Wilhelmy plates, the simulated lateral pressure is spatially averaged (equivalently, integrated and normalized) over the x-direction of each sensor.
Boundary conditions arising from the 2D channel geometry are implemented numerically and account for both the outer trough and the inner channel in which wave propagation occurs.
At the outer boundaries of the trough, we impose hard no-slip conditions that constrain the finite experimental domain.
At the inner channel walls, we impose full tangential slip to reflect the hydrophobic Teflon surface, while enforcing strict no-penetration (no-slip in the normal direction).

\section{Results and Discussion}

\subsection{Steady state isomerization}

In order to better understand how isomerization of azoPC can excite mechanical pulses via pressure changes, 
isotherms (surface pressure as a function of area) have been measured for the all-trans and all-cis states 
(achieved by 30 minute illumination with 455/365nm light). 

\begin{figure}
  \centering
  \includegraphics[scale = 1]{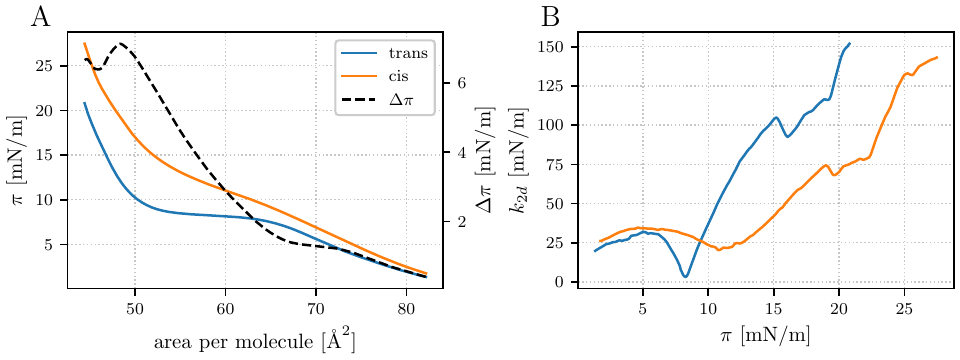}
  \caption{A: Pressure-area isotherms of a monolayer consisting of 80:20 mol\% DPPC:azoPC at a temperature of  $20\pm 0.5 {\rm ^\circ C}$ in all-trans and all-cis isomerization states, achieved by 30 minutes of illumination with 365\,nm (cis)/455\,nm (trans) light and their lateral pressure difference $\Delta\pi = \pi_{cis}-\pi_{trans}$. Lateral pressure is strictly higher in the all-cis state, but the magnitude of this effect is state dependent. B: The corresponding isothermal compression modulus $k_{2d}=1/\kappa_T$ as a function of pressure. Local minima correspond to the LE/LC phase transition, for which the monolayer is very soft. Trans to cis isomerization reduces the prominence of the minimum while increasing the pressure at which it occurs.
}
  \label{fgr:isotherm}
\end{figure}

The resulting state diagrams are plotted in Fig.\ \ref{fgr:isotherm}A. Although lateral pressure is strictly higher for the cis state, the difference is dependent on state (Fig.\ \ref{fgr:isotherm}B), rising slowly throughout the liquid expanded (LE) phase and faster throughout the phase transition (PT) until reaching a maximum at a molecular area of 
$50 \angstrom^2$, close to where the liquid condensed (LC) / solid ordered (SO) transition is expected.
Further compression yields a decrease in pressure difference.
As seen in Fig.~\ref{fgr:isotherm}B, isothermal compression moduli also vary greatly between isomerization states.
For trans azoPC the profile closely resembles a pure DPPC isotherm with a prominent minimum at the phase transition, whereas for cis azoPC the minimum is less prominent and shifted toward a higher lateral pressure.
AzoPC is in a liquid expanded state \cite{Pernpeintner2017} and does not undergo a phase transition at room temperature.
The influence on the transition is therefore because cis-azoPC occupies a larger area per molecule and thus disrupts the packing of side chains to form the ordered phase.
This effect is similar to introducing unsaturated phospholipids into the monolayer \cite{Luviano2019}.
Excitation then is the result of local switching between the ground state (all-trans for pulses of compression and all-cis for pulses of expansion) and an intermediate between the two isotherms shown in Fig.~ \ref{fgr:isotherm}A.
Since the trough area is held constant during pulses, the difference between the steady states before and after the pulse is isochoric.

\subsection{Pulse measurements}

\subsubsection{Unconstrained pulses}

\begin{figure}
  \centering
  \includegraphics{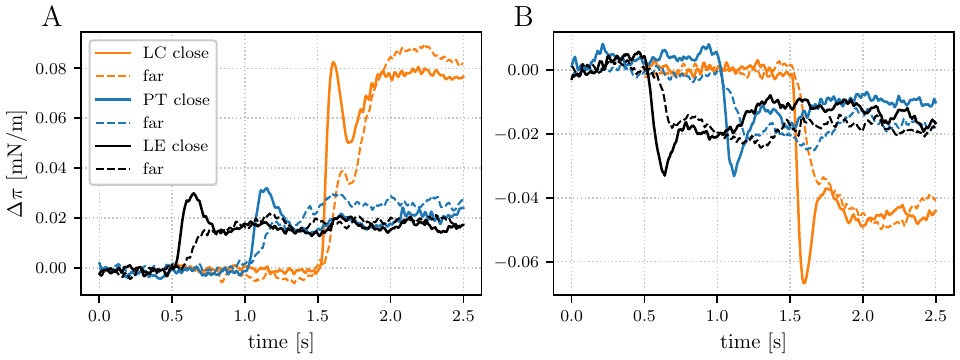}
  \caption{A: Trans-to-cis isomerization yields compression pulses on an unconstrained monolayer.
  Pulses are very similar between the LE phase and the phase transition (PT), whereas both amplitude and pulse duration change significantly for the LC phase. B: Cis-to-trans isomerization was achieved by first illuminating the monolayer with UV light for 30 minutes and using visible excitation light.
  Pulse morphology only differs by the sign of the pressure change, indicating no preference for either isomerization reaction.
  Measurements were performed on a monolayer consisting of 80:20 mol\% DPPC:azoPC at $20\pm 0.5 {\rm ^\circ C}$. Pulse data is averaged over 10 measurements and amplitude is calculated by $\Delta\pi (t) = \pi(t)-\pi_0$ with $\pi_0 = 5\,\mathrm{mN/m}$ for LE, $\pi_0 = 8\,\mathrm{mN/m}$ for PT and $\pi_0 = 12\,\mathrm{mN/m}$ for LC. Pulses are staggered by 0.5s for better readability.}
  \label{fgr:pulses_free}
\end{figure}

Figure\ \ref{fgr:pulses_free} shows pulses of compression (A) and expansion (B) in LE and LC phases, as well as in the transition regime in the absence of any constraining channels.
Pulses were excited by illuminating a circle with 5cm diameter next to the two sensors, which were 3cm apart.
Since no constraining barriers were used, the perturbation spreads in all directions thus yielding only low amplitudes.

All measurements from sensor 1 show a transient spike in pressure, followed by a rarefaction and small oscillations before settling to the new steady state pressure. This pulse quickly dissipates as evidenced by the second sensor, where the pressure signal resembles a smeared out version of the quasi-discontinuous increase in pressure which the rapid photoswitching induces.
For LC measurements a strongly attenuated pulse is observed as a kink in the rising flank of the second sensor.
As expected from Fig.~\ref{fgr:isotherm}B, the increase/decrease in pressure is larger in the LC phase, whereas contrary to the steady state considerations, no measurable difference is observed between the LE and phase transition regime.
Overall the differences in steady state pressures is small compared to the maximum difference in Fig~\ref{fgr:isotherm}, so the induced change to the state diagram is also small, 
and all-trans/cis isothermal can be used for simulation.
A major difference between pulses of compression and expansion is not observed. 

\begin{figure}
    \centering
    \includegraphics{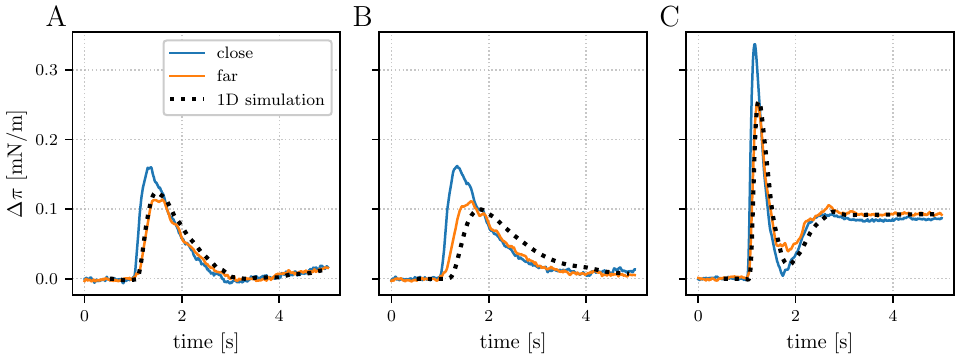}
    \caption{Adding a confining channel (width: 3cm, length: 15cm) which is closed on the site of excitation greatly increases amplitude, timescale and stability of the measured pulses, while the differences between LC (C) phase and LE/PT (A, B) become even more pronounced.
    1D simulations according to eq.~\eqref{eq:DGL}, with close signal as input, match experimental data very well in LE (A) $k_{sim}=0.030\,\mathrm{mN/m}$ and LC  (C) phase $k_{sim}=0.075\,\mathrm{mN/m}$, but predict a slower pulse in the PT (B) regime $k_{sim}=0.005\,\mathrm{mN/m}$.
    Measurements were performed on a monolayer consisting of 80:20 mol\% DPPC:azoPC at $20\pm 0.5 {\rm ^\circ C}$. Pulse data is averaged over 10 measurements and amplitude is calculated by $\Delta\pi (t) = \pi(t)-\pi_0$ with $\pi_0 = 5\,\mathrm{mN/m}$ for LE, $\pi_0 = 8\,\mathrm{mN/m}$ for PT and $\pi_0 = 12\,\mathrm{mN/m}$ for LC.}
    \label{fgr:pulses_channel}
\end{figure}

\begin{figure}
    \centering
    \includegraphics{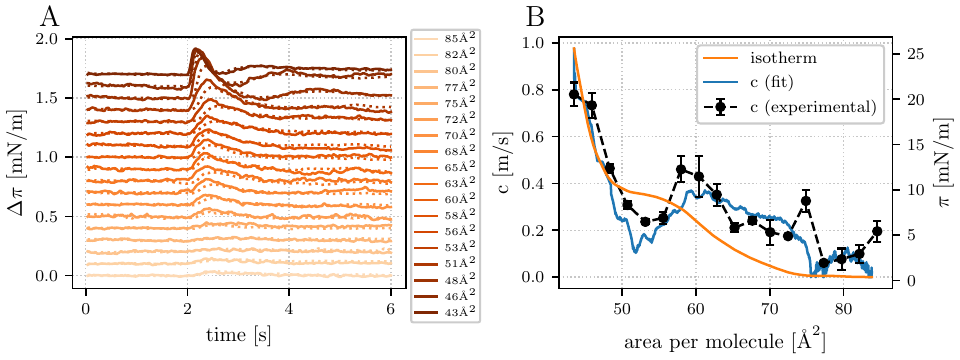}
    \caption{A: Compression pulses in a channel (width: 3cm, length: 15cm) measured at equally spaced areas per molecule along an isotherm consisting of 80:20 mol\% DPPC:azoPC at $21\pm 0.5 {\rm ^\circ C}$.
    Pulse data is averaged over 3 measurements and data is offset along the y-axis for better visibility.
    Dashed lines represent signals from the far sensor.
    B: Pulse velocities calculated by cross correlation of the close and far signals for the pulses displayed in A
    are fitted with the propagation velocity predicted by eq.\ \eqref{eq:phase} with $k_{2d}$-values derived from the isotherm and $\omega = (6.37\pm 1.76)\,\mathrm{s}^{-1}$ as the only fit parameter.
    }
    \label{fgr:velocity}
\end{figure}

\subsubsection{Pulses in channels} 

To limit dissipation to one dimension, a PTFE channel, which is closed on the site of excitation and has both sensors inside of it, has been used.
Results are plotted in Fig.~\ref{fgr:pulses_channel} and Fig.~\ref{fgr:velocity}A 
for a channel with a length of 15cm and a width of 3cm.
The reversible aspects of the perturbation, i.e. the pulse, now greatly outweighs the change in steady state pressure.
This increase in amplitude is accompanied by an elongation of the pulses from a maximum length of 300\,ms for the unconstrained case up to 2\,s with barriers.
This also amplifies the variations between the measured phases, where the LC phase displays a much shorter pulse of higher amplitude with a prominent rarefaction.
Simulations based on the 1D model given by eq.~\eqref{eq:DGL} match experimental data very well, thus confirming that the channel has the desired effect of reducing the dimension of the problem, see Fig.~\ref{fgr:pulses_channel}.
Only in the phase transition (Fig.~\ref{fgr:pulses_channel}) a visible difference in propagation velocity is observed, which is likely due to the difference in elastic properties between the isothermal steady-state measurement used for simulation and the adiabatic nature of the pressure pulse.
The overall similarity of pulses in the LE phase and PT regime is expected for lipid monolayers, whose elastic properties in the PT closely resemble the LE phase due to dynamic stiffening \cite{Arriaga2010}.

Further investigation of pulse velocity is displayed in Fig.~\ref{fgr:velocity}, where pulses were measured along an isotherm to confirm the state dependence of propagation velocity predicted by eq.~\eqref{eq:phase}.
This equation is valid for frequencies $\omega \ll 10^{12}\,1/s$, and predicts a much lower velocity than the well-known relation $c = \sqrt{k_{2d}/\rho}$ due to viscous coupling to the subphase.
As is displayed in Fig.~\ref{fgr:velocity}, a good fit to experimental data is achieved with $\omega = (6.37\pm 1.76)\,\mathrm{s}^{-1}$ as the only fit parameter.
Since eq.~\eqref{eq:phase} is the phase velocity, the calculated frequency does not have a defined physical meaning, but it is in agreement with Fourier spectra of pulses, which are dominated by low frequencies, and yields wavelengths greater than the diameter of the excited region.
The frequency is also compatible with the 
response time $\tau \simeq 100\,\mathrm{ms}$ (i.e., $\omega \lesssim 1/\tau$) 
obtained from the 2D simulations, which will be discussed later on.
Other methods of pulse excitation have verified these predictions as well using frequencies up to 1\,Hz \cite{Griesbauer2012,Kappler2017}. 
Since our fit used the isothermal modulus ($k_{2d}=1/\kappa_T$) which is smaller than the adiabatic one, higher velocities would be expected for an adiabatic phenomenon, which would then necessitate choosing a lower frequency.

In order to further investigate how channel geometry influences pulse measurements, different channel lengths have been used at fixed channel width in Fig.~\ref{fgr:pulse3}  and different channel width at 
fixed channel length in Fig.~\ref{fgr:pulse4}. In the following we will discuss all experimental results 
regarding the pulse shapes in comparison to 1D and 2D simulations.

\subsection{1D simulations: pulse propagation and nonlinearities}

First, we compare the experimental results for pulse propagation in a narrow channel to 1D simulations based on the 1D model \eqref{eq:DGL} in Fig.~\ref{fgr:pulses_channel}.
 For this comparison, experimental data from the first (close) pressure sensor were used to generate
 appropriate time-dependent boundary conditions for the 1D model. 
 Then the experimental data from the second (far) pressure sensor is compared to the resulting pressure 
 simulation data. 
 This procedure focuses on the question whether the 1D fractional wave equation \eqref{eq:DGL}
 correctly captures the propagation of a pulse, i.e., the evolution of the pressure pulse shape from the first to the second sensor.  The procedure minimizes uncertainties associated with the details of pulse generation and does not require a model for pulse generation 
 because experimental data at the first sensor is directly used as boundary condition. 
 These details involve  the kinetics of the 
 isomerization process of  azoPC and the resulting surface pressure 
 response times after isomerization, which will be addressed in more detail below in the framework of the 2D model. Consequently these 1D simulations of pulse propagation contain \emph{no} fitting parameters
 describing the pulse generation process.
 
 In general, we find quantitative agreement between experimental and simulated surface pressure pulses at the second sensor for different surface pressure levels $\pi_0$ (i.e., different lipid phases along the isotherm), as shown in Fig.~\ref{fgr:pulses_channel}.
 These simulations were conducted by directly taking all material parameters directly from measurements, i.e. the isotherm, which completely renounces the use of fitting parameters. 

\begin{figure}
    \centering
    \includegraphics[scale = 1 ]{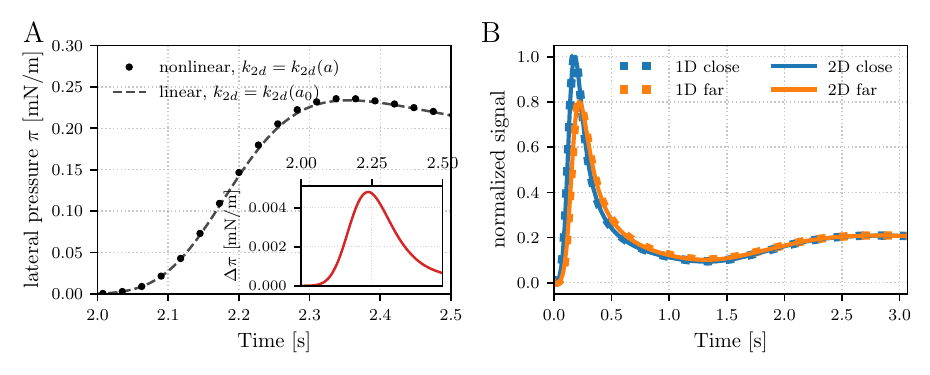}
    \caption{
    A: Comparison between two numerical solutions of eq.~\eqref{eq:DGL} obtained with a linear elastic modulus (dashed) and with a state-dependent (nonlinear) modulus (dots). The boundary condition is obtained  from the experimental pressure trace $\pi_1(t)=\pi_0 + \Delta \pi_1(t)$ at the first sensor 
    at a base pressure level $\pi_0 = 17.5\,\mathrm{mN\,m^{-1}}$, while the elastic modulus is taken from the corresponding isotherm as explained in section \emph{Simulation}.
    The inset shows the difference between both simulations, indicating that isotherm-induced nonlinearities have a negligible influence under the conditions considered.
    B:
    Convergence between the one- and two-dimensional models, eqs.~\eqref{eq:DGL} and \eqref{eq:DGL2d}, for a long and narrow channel ($0.3 \, \mathrm{m}$ length, $0.02 \, \mathrm{m}$ width) and for a Gaussian-shaped excitation (using identical material parameters in both simulation).
    }
    \label{fig:1d_results}
\end{figure}

In  simulations, we can also explore the influence of nonlinearities via the 
area-dependence $k_{2d}$ of the compression modulus in a controlled fashion, 
see Fig.\ \ref{fig:1d_results}A. 
For a purely linear equation, we employ $k_{2d}=k_{2d}(a_0)$ with the fixed average area in the 
1D model \eqref{eq:DGL} (dashed line in Fig.\ \ref{fig:1d_results}A).
The full nonlinear model including the dependence $k_{2d}=k_{2d}(a)$ on the local area 
$a = a_0\!\left(1+\partial_x u\right)$ (dots in Fig.\ \ref{fig:1d_results}A) 
shows little to no discernible influence of this type of nonlinearity.
One reason that nonlinearities do not play a role is the small amplitude 
of pressure pulses that are generated by azoPC photoswitching. 
For the data in Fig.\ \ref{fig:1d_results}A the pulse amplitude is $\Delta \pi <0.25 \,\mathrm{mN/m}$ while the base pressure is $\pi_0 = 17.5\, \mathrm{mN/m}$, such that the relative amplitude is only $\Delta \pi/\pi_0 < 2\%$.

According to our above scaling analysis we expect non-negligible nonlinear 
effects for  large displacement gradients or strains  $U/L \sim 1$;
this can also be seen directly from the isothermal elastic modulus $k_{2d}(a)$, 
where the area per lipid $a = a_0 \left( 1 + \partial_x u \right)$
will be left unchanged if $\partial_x u = \epsilon \ll 1$.
In simulations we can employ excitation amplitudes that are much larger than what can be realized by 
photoswitching and find nonlinear effects for strains $\epsilon \gtrsim 0.2$ confirming these expectations
(see Supporting Information).

\subsection{2D simulations: effects from channel geometry}

Geometric effects from different channel geometries provide a sensitive test of the 
theory based on the fractional wave equation, which requires to move from 1D simulations to full 2D simulations.

\begin{figure}
    \centering
    \includegraphics[scale = 0.8]{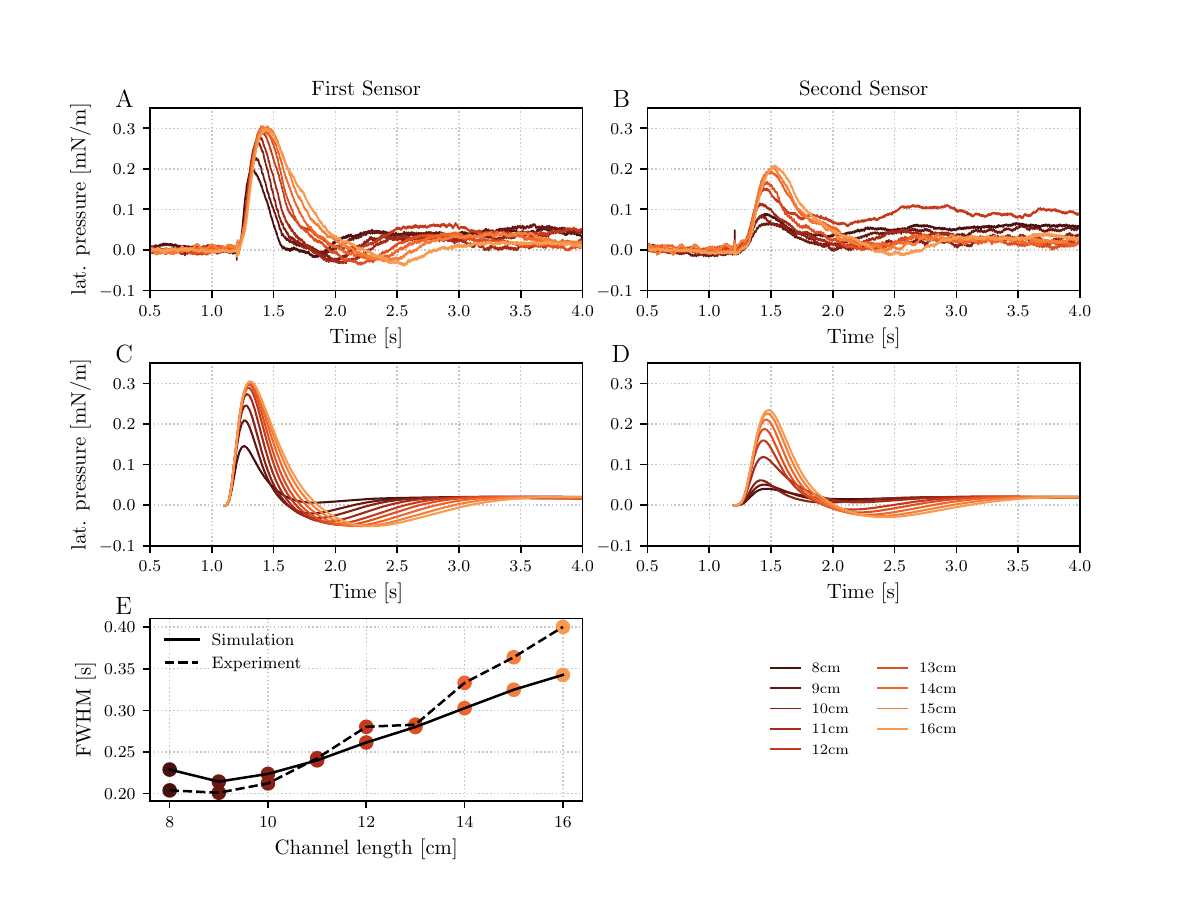}
    \caption{
    Experimental results (A,B) for different channel lengths, constant channel width of $1.5 \, \mathrm{cm}$ and the corresponding simulations (C,D) obtained from the two dimensional framework, given by eq.~\eqref{eq:DGL2d}. Calculations from both the experiment and the simulations of the full width at half maximum (FWHM) can bee seen in E.
    In the simulations, we retain only the dilatational contributions, using a constant dilatational elasticity of $\lambda_s = 0.05\,\mathrm{N\,m^{-1}}$ and a constant dilatational viscosity of $\zeta_s = 1\times 10^{-4}\,\mathrm{N\,s\,m^{-1}}$ for all channel lengths. }
    \label{fgr:pulse3}
\end{figure}

\begin{figure}
    \centering
    \includegraphics[scale = 0.8]{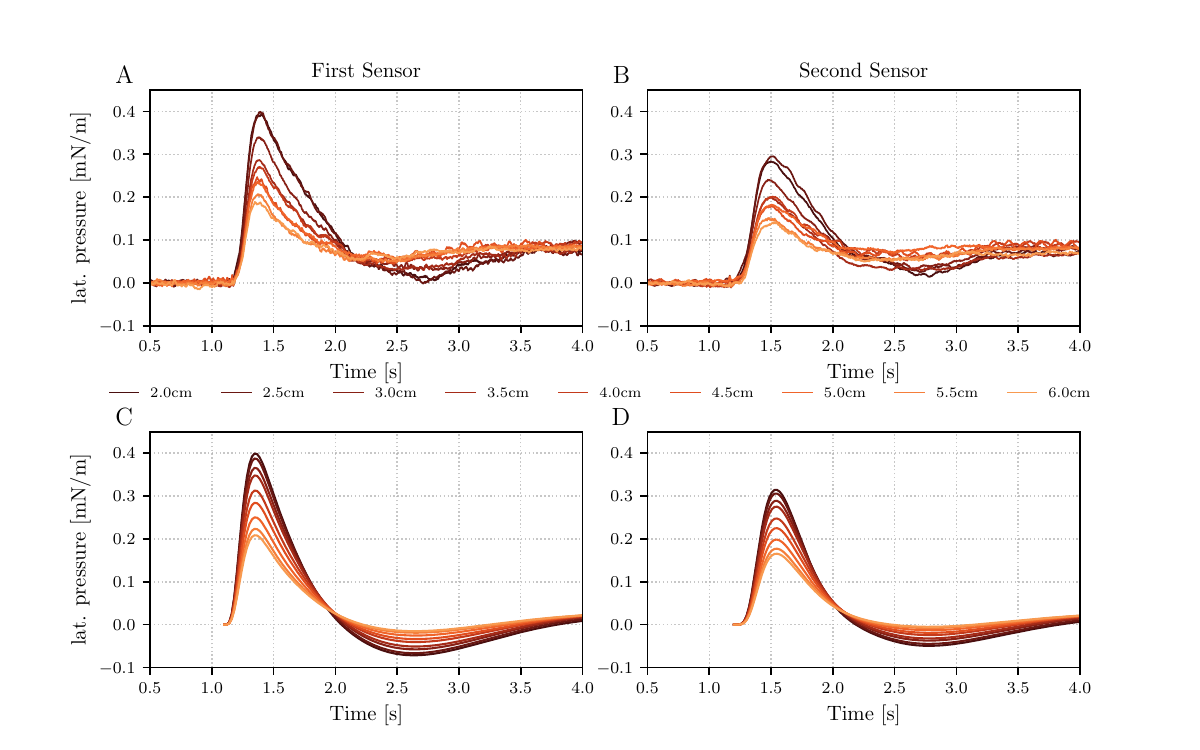}
    \caption{
    Experimental results (A,B) for different channel widths, constant channel length of $17 \, \mathrm{cm}$ and the corresponding simulations (C,D) obtained from eq.~\eqref{eq:DGL2d}.
    In the simulations, we retain only the dilatational contributions, using a constant dilatational elasticity of $\lambda_s = 0.05\,\mathrm{N\,m^{-1}}$ and a constant dilatational viscosity of $\zeta_s = 1\times 10^{-4}\,\mathrm{N\,s\,m^{-1}}$ for all channel lengths. }
    \label{fgr:pulse4}
\end{figure}

In Fig.\ \ref{fig:1d_results}B, we first show that for narrow and long channels 2D simulation results agree with the 1D simulations as expected. 
In two dimensions, our simulations reproduce the experimentally observed 
geometric effects imposed by different channel designs, a constant width of $1.5\,\mathrm{cm}$ with varying channel length (Fig.~\ref{fgr:pulse3}) or a constant channel length of $17\,\mathrm{cm}$ with varying channel width (Fig.~\ref{fgr:pulse4}).
In particular, an early plateau in the wide-channel configuration and a comparatively soft overshoot in the narrow-channel configuration clearly distinguish the two responses, despite the fact that the same governing equation is employed in both cases.

In  the 2D simulations we include the excitation process into the model via the forcing term, see eq.~\eqref{eq:f_t}. 
We use two parameters of the excitation process, 
the characteristic response time $\tau$ for the surface pressure change after isomerization and the spatial width $\xi$ of the Gaussian excitation profile as fit parameters 
to match the experimental pulse shapes using \emph{one} set of fit parameters for all shapes from Figs.~\ref{fgr:pulse3} and \ref{fgr:pulse4}. 
We employ a differential evolution algorithm \cite{Storn1997} to minimize the mean square deviations 
between  experimental and simulated pulse shapes to determine the fit parameters and find 
 a spatial width $2\xi~=~4 \, { \rm cm }$ comparable to the diameter of $5cm$ 
 of the illuminated circle in the experiments  and 
and a characteristic response time  $\tau  \simeq 100\, \mathrm{ms}$. 
The  parameter $\tau$ to fit the experimental data on pulse shapes
also provides insights into the kinetics of the photoinduced switching of the monolayer.

It is well-known that the isomerization process of a single azobenzene group after photon absorption  is very fast (in the ps regime  \cite{Satzger2003,Bandara2012}). The switching time of a macroscopic ensemble of azobenzene containing surfactants in solution 
can reach the second regime and decrease with light intensity \cite{Stranius2017,Arya2020} because only a fraction of the 
surfactants absorb photons at any moment. If these surfactants form micelles, the switching kinetics is slowed down \cite{Arya2020}. 
When azoPC is embedded in a monolayer,  
the kinetics of the resulting surface pressure change is also much slower \cite{Backus2011,Arya2020,Warias2023} with switching times  in the 
second regime.   In  red blood cell membranes, 
comparable slow surface pressure response times of  $200-500 \mathrm{ms}$ have been reported \cite{Hglsperger2023} 
(decreasing with light intensity). These results show that steric effects in the molecular environment can  
obstruct the isomerization of the  azobenzene group \cite{Pritzl2025,Arya2020}.
In more complex environments, such as monolayers at an air-water interface or membranes, 
additional relaxation kinetics within the monolayer or membrane are required after photoisomerization of azoPC 
before a  surface pressure change occurs. 
This relaxation involves local rearrangement of lipid tails \cite{Backus2011,Warias2023} and,  for mechanical 
relaxation, also motion 
of lipids within the monolayer against viscous and hydrodynamic forces. 
 Therefore, the characteristic time $\tau$ in the pulse excitation process
 should not be interpreted as the molecular photoisomerization time of azobenzene, which is much faster, but as the effective monolayer pressure response time after optical excitation. This effective timescale includes the finite photon flux of the flash, photoswitching kinetics in the packed monolayer, steric constraints from lipid packing, local rearrangement of lipid tails, and mechanical relaxation of the monolayer and is much slower. 
To further constrain the physical meaning of the response time, we directly 
placed a Wilhelmy plate at the center of the illuminated region. 
These measurements directly confirm a characteristic initial surface-pressure response time of $\tau \simeq 100\, \mathrm{ms}$
(see Supporting Information). 
Experiments with different flash light intensities show that decreasing the light intensity only decreases the amplitude of pressure pulses but leaves the 
response time $\tau$ unaffected (see Supporting Information). This shows that the surface pressure response time is 
not limited by light intensity (the available photons) but by the rearrangement and relaxation kinetics following the 
photoisomerization.
The slowest process contributing to $\tau$ is mechanical relaxation by lipid motion 
within the monolayer against viscous and hydrodynamic forces. This mechanical relaxation is exactly captured by the 
equation of motion  \eqref{eq:DGL}. 
Therefore, the scaling (\ref{eq:scaling}) implied by the dispersion relation \eqref{eq:disperion} and discussed above
can be used to estimate $\tau$ for an excitation on the scale $L=2\xi~=~4 \, \mathrm{cm}$
in the relevant limit where
hydrodynamic friction via the subphase dominates over monolayer viscosity.
For $L=2\xi~=~4 \, \mathrm{cm}$,  we find $\tau \sim 50-150\,\mathrm{ms}$ further corroborating our fit result. 
We also note that timescales $\tau \sim 50-150\,\mathrm{ms}$ are compatible with the fit result
$\omega = (6.37\pm 1.76)\,\mathrm{s}^{-1}$ obtained in Fig.\ \ref{fgr:velocity}B in the sense that $1/\omega \lesssim \tau$.

The forcing amplitude $f_0$ in eq.~\eqref{eq:f_t} is not determined by fitting the entire 
pulse shape but from the maximal pressure in a pulse:
We always simulate \eqref{eq:DGL2d} for a unit amplitude forcing $\bar{f}_0=1$ 
and determine $f_0$ by comparing the experimental maximum pressure value 
$\pi_{\mathrm{exp,max}}$ of a pulse to the 
numerically obtained maximum pressure value $\bar{\pi}_{\mathrm{num,max}}$
by $f_0 = \pi_{\mathrm{exp,max}} / \bar{\pi}_{\mathrm{num,max}}$
(neglecting the influence of nonlinearities and exploiting the approximative linearity of  \eqref{eq:DGL2d}). We obtain 
$f_0 \approx 0.1\,\mathrm{mN/m^2}$.

\section{Conclusions}

Our experimental and theoretical results clearly demonstrate that rapid photoswitching is a viable method of exciting two-dimensional longitudinal acoustic pulses in lipid monolayers in a controlled and reproducible fashion, opening up possibilities to investigate more complex membranes and living systems. 
Pulses can be quantitatively described by a nonlinear fractional wave equation (eqs.\ \eqref{eq:DGL} or \eqref{eq:DGL2d}), where a term containing a  fractional time derivative of order $3/2$ captures the hydrodynamics of the monolayer subphase, while  the wave equation can be formulated purely in a surface displacement field of the monolayer.
We find that a 1D equation (\ref{eq:DGL}) describes pulse propagation along a narrow channel quantitatively in good agreement with experiments \emph{without} any fit parameters. A 2D equation (\ref{eq:DGL2d}) also captures effects arising from the channel geometry correctly for wider channels.
Generation of pulses by photoisomerization of azoPC can be included into the 
equation by a forcing term containing a characteristic response time $\tau$ of the monolayer. We obtain good agreement between simulation and experimental pulse shapes for a response time $\tau \simeq 100\, \mathrm{ms}$. 

This characteristic response time for surface pressure changes
 is set by the slow relaxation via lipid motion 
within the monolayer against viscous forces from hydrodynamic 
coupling to the subphase.

The fractional  wave equations are nonlinear via an area-dependent compression modulus. In simulations we only find negligible  influence of the nonlinearity
on the pulse shapes. A possible reason for this observation is the small amplitude 
of pressure pulses that are generated by azoPC photoswitching (below $0.25 \,\mathrm{mN/m}$). 
Photoswitching typically generates pulses with a relative amplitude of only $2\%$ compared to the base pressure of the monolayer, even  in narrow channel geometries.  

While the photoisomerization of azobenzene has been explored in great depth \cite{Pritzl2025}, its effect on membranes as part of a phospholipid is still a topic of recent research \cite{Socrier2023}.
On a molecular level, excitation is a consequence of a local increase in area, while the macroscopic effects are an increase in pressure and consequently a change in material properties.
The monolayer is kept at equilibrium during excitation, but the local state in the illuminated area changes on the order of 100\,ms, resulting in an adiabatic process.
The measured pulses expose the viscoelastic nature of the monolayer and the hydrodynamic coupling to the viscous subphase, since the local step-like increase in molecular area neither leads to a step-like increase in lateral pressure (purely elastic), nor does it lead to a completely transient spike (purely viscous), but a superposition of both.
The channel measurements show how pressure relaxation is limited by the characteristic speed of sound in the monolayer and how geometry influences the morphology and stability of pulses, which is to be expected for acoustic phenomena.
Temporal pulse width is a function of both channel length and propagation velocity, which would mean that the length of the unconstrained pulse represents the fastest relaxation possible for a given excitation strength and monolayer state.

Comparison to previous methods of pulse excitation clearly demonstrate advantages and limitations of rapid photoswitching.
Pulses excited by addition of solvent \cite{Griesbauer2012,Shrivastava2013} display very irregular pulse shapes and have to be spatially separated from the site of measurement to limit the influence of capillary waves, thus decreasing the measured amplitude to a level comparable to our results, whereas excitation by short bursts of gas \cite{Fichtl2016,Fichtl2016b} yields higher pulse amplitudes (up to $2.5 \,\mathrm{mN/m}$, i.e., one order of magnitude larger than for photoswitching) and more consistent pulse shape.
Both methods produce pulses with temporal widths on the order of seconds, which, following from our results for unconstrained layers, indicates that the exciting perturbation persists longer than the monolayers relaxation time.
In contrast to that, pulses excited with a piezo-actuated cantilever display timescales on the order of milliseconds and nonlinear behavior close to the phase transition as predicted by Heimburg and Jackson \cite{Heimburg2005}.
A direct comparison to our results is not possible, since FRET was used as a detection mechanism instead of lateral pressure measurement but the difference in timescale and its sensitivity to the phase transition can be discussed qualitatively.
As outlined earlier, the timescale of photoisomerization is influenced by the viscous environment of the azobenzene and leads to response times of $\tau \simeq 100\,\mathrm{ms}$ in our experiments, thus rendering it impossible to excite faster dynamics.
Mechanical actuation, however, does not have these constraints making a higher rate of compression achievable, thus enabling nonlinear behavior.
The next step now is to apply our methods, both experimental and theoretical, to more complex systems to further investigate acoustic phenomena in biological membranes, such as interactions with enzymes and chemicals, the influence of the rigidity and viscosity of the subphase and ultimately the influence on living systems.

\section*{Acknowledgements}

This work was funded by the Deutsche Forschungsgemeinschaft (DFG, German Research Foundation)- project numbers 387637964 and 539638202, as well as TU Dortmund University.

\section*{Supporting information}

We present  supplementary measurements regarding the pressure pulse  excitation and its dependence on light intensity as well as additional simulations illustrating the influence of the nonlinearities for large excitation amplitudes and confirming 
the scaling argument that nonlinearities will play a role if typical displacement gradients $U/L$ are sufficiently large.

\bibliography{lit.bib}

\end{document}